\documentclass[reprint,12pt]{elsarticle}
\usepackage{amsmath}
\usepackage{lineno,hyperref}
\newcommand{\bes}{\begin{subequations}}
\newcommand{\ees}{\end{subequations}}
\newcommand{\bea}{\begin{eqnarray}}
\newcommand{\eea}{\end{eqnarray}}
\newcommand{\ben}{\begin{equation}}
\newcommand{\een}{\end{equation}}
\newcommand{\al}{\alpha}
\newcommand{\ba}{\beta}
\newcommand{\del}{\delta}
\newcommand{\Del}{\Delta}
\newcommand{\ga}{\gamma}
\newcommand{\sech}{\mbox{ sech}}
\journal{Physics Letters A}

\begin{document}

\begin{frontmatter}

\title{Nonstandard bilinearization of $\cal{PT}$-invariant nonlocal nonlinear Schr\"{o}dinger equation: \\Bright soliton solutions}

\author[bdu]{S.~Stalin}
\author[bdu]{M.~Senthilvelan\corref{auth1}}
\address[bdu]{Centre for Nonlinear Dynamics, School of Physics, Bharathidasan University, Tiruchirappalli 620024, Tamil Nadu, India}
\cortext[auth1]{Corresponding author; Phone : +91 431 2407057; Fax : +91 431 2407093}
\ead{velan@cnld.bdu.ac.in}
\author[bdu]{M.~Lakshmanan}

\begin{abstract}
In this paper, we succeed to bilinearize the $\cal{PT}$-invariant nonlocal nonlinear Schr\"{o}dinger (NNLS) equation through a nonstandard procedure and present more general bright soliton solutions. We achieve this by bilinearizing both the NNLS equation and its associated parity transformed complex conjugate equation in a novel way. The obtained one and two soliton solutions are invariant under combined space and time reversal transformations and are more general than the known ones. Further, by considering the two-soliton solution  we bring out certain novel interaction properties of the $\cal{PT}$-invariant multi-soliton solutions.
\end{abstract}


\end{frontmatter}

%\linenumbers

\section{Introduction}
 In this work, we intend to develop a nonstandard bilinearization procedure to generate more general bright soliton solutions for the NNLS equation \cite{1}, namely 
\begin{equation}
iq_{t}(x,t)-q_{xx}(x,t)-2q(x,t)q^{*}(-x,t) q(x,t)=0,
\label{1.1}
\end{equation}
where $q(x,t)$ is a complex valued function and $q_t$ and $q_{x}$ represent the derivatives of $q$ with respect to $t$ and $x$, respectively. In the above equation, $q^{*}(-x,t)$ is the nonlocal term which implies that the solution $q(x,t)$ of (\ref{1.1}) evaluated at $x$ requires information from $-x$ also \cite{2}. Eq. (\ref{1.1}) is obviously $\cal{PT}$-symmetric \cite{3}. As proved by Ablowitz and Musslimani recently, Eq. (\ref{1.1}) admits a Lax pair and infinite number of conserved quantities \cite{4} and hence it is a completely integrable nonlinear evolution equation \cite{5}. To the best of our knowledge, a correct bilinear formalism for the NNLS is to be yet reported in the literature. In the following we present a novel bilinearization procedure to derive general bright soliton solution for the class of NNLS equations of the type (\ref{1.1}). To obtain such soliton solutions of (\ref{1.1}) through Hirota's bilinearization process, we also augment the parity transformed complex conjugate equation of (\ref{1.1}) because we treat the nonlocal term $q^{*}(-x,t)$ as  a separate quantity. The parity transformed complex conjugate equation of (\ref{1.1}) reads
\bea
iq_{t}^{*}(-x,t)+q_{xx}^{*}(-x,t)+2q^{*}(-x,t)q(x,t) q^{*}(-x,t)=0.
\label{1.2}
\eea 
We bilinearize Eqs. (\ref{1.1}) and (\ref{1.2}) simultaneously by introducing the following dependent variable transformations, namely
\ben
q(x,t)=\frac{g(x,t)}{f(x,t)}, \quad q^{*}(-x,t)=\frac{g^{*}(-x,t)}{f^{*}(-x,t)},
\label{1.4} 
\een
where $g(x,t)$ and $g^{*}(-x,t)$ are complex functions while $f(x,t)$ and $f^{*}(-x,t)$ are also in general complex functions and all of them are distinct. To obtain the correct bilinear equations, we adopt a non-standard method by introducing appropriate number of auxiliary functions. This procedure is similar to the technique followed in \cite{6} in the case of Sasa-Satsuma higher-order NLS equation and in \cite{7} in the case of coherently coupled NLS equation. From these two studies one can find that the introduction of suitable number of auxiliary functions in the bilinear process leads to match the number of bilinear equations to be equal to the number of unknown functions, which leads to the nontrivial solutions. In the present NNLS equation, by applying the transformations for $q(x,t)$ and $q^*(-x,t)$ one finds that there are certain terms which are not in bilinear forms. To make them bilinear one can introduce appropriate auxiliary variables and then proceed with the analysis in the standard way. By adopting this procedure, we introduce four auxiliary functions in the bilinear process of NNLS equation. This helps us to match the number of unknown functions with the bilinear equations. Through this procedure we explore more general soliton solutions for the NNLS equation as described below. 
Substituting (\ref{1.4}) into (\ref{1.1}) and (\ref{1.2}) gives the following bilinear equations, namely \bes
\bea
&&D_1~g(x,t)\cdot f(x,t)=-2g^{*}(-x,t) \cdot s^{(1)}(-x,t),\label{1.5} \\
&&D_2~g^{*}(-x,t)\cdot f^{*}(-x,t)=2g(x,t) \cdot s^{(2)}(-x,t), \\
&& D_3~f(x,t)\cdot f(x,t)=4h^{(1)}(-x,t) \cdot f(x,t),\\
&&D_3~f^{*}(-x,t)\cdot f^{*}(-x,t)=4h^{(2)}(-x,t) \cdot f^{*}(-x,t),
\label{1.6}
\eea \ees
where $D_1\equiv(iD_t-D_x^2)$, $D_2\equiv(iD_t+D_x^2)$, $D_3\equiv D_x^2$ and the auxiliary functions are defined by the additional set of following four bilinear equations:
\bes
\bea
s^{(1)}(-x,t)\cdot f^{*}(-x,t)=g^{2}(x,t),~ h^{(1)}(-x,t) \cdot f^{*}(-x,t)=g(x,t) \cdot g^{*}(-x,t)\\
s^{(2)}(-x,t) \cdot f(x,t)=g^{*2}(-x,t),~ h^{(2)}(-x,t) \cdot f(x,t)=g(x,t) \cdot g^{*}(-x,t).
\label{1.7}
\eea \ees
In the above $D_{t}$ and $D_{x}$ are the standard Hirota's bilinear operators \cite{8} and they are defined as,
\bea
D_x^{m}D_t^{n}(g\cdot f) =\bigg(\frac{\partial}{\partial x}-\frac{\partial}{\partial x'}\bigg)^m\bigg(\frac{\partial}{\partial t}-\frac{\partial}{\partial t'}\bigg)^n g(x,t)\cdot f(x',t')\Big|_{ (x=x', t=t')}.\nonumber
\eea
Soliton solutions can be obtained by solving the above set of bilinear equations (\ref{1.5})-(\ref{1.6}). We expand the unknown functions $g(x,t)$, $g^{*}(-x,t)$, $f(x,t)$, $f^{*}(-x,t)$, $s^{(1)}(-x,t)$, $s^{(2)}(-x,t)$, $h^{(1)}(-x,t)$ and $h^{(2)}(-x,t)$ in terms of a small parameter $\epsilon$, that is
\bes
\bea
\hspace{-1.0cm}&&g(x,t)=\epsilon g_{1}+\epsilon^{3} g_{3}+... ~~~~~~~~~~~~ g^{*}(-x,t)=\epsilon g_{1}^{*}+\epsilon^{3} g_{3}^{*}+...\\
\hspace{-1.0cm}&&f(x,t)=1+\epsilon^{2} f_{2}+\epsilon^{4} f_{4}+...~~~~~ f^{*}(-x,t)=1+\epsilon^{2}f^{*}_{2}+\epsilon^{4}f^{*}_{4}+...\\
\hspace{-1.0cm}&&s^{(1)}(-x,t)=\epsilon^{2}s^{(1)}_{2}+\epsilon^{4}s^{(1)}_{4}+...~~~ s^{(2)}(-x,t)=\epsilon^{2}s^{(2)}_{2}+\epsilon^{4}s^{(2)}_{4}+...\\\
\hspace{-1.0cm}&&h^{(1)}(-x,t)=\epsilon^{2}h^{(1)}_{2}+\epsilon^{4}h^{(1)}_{4}+...~~h^{(2)}(-x,t)=\epsilon^{2}h^{(2)}_{2}+\epsilon^{4}h^{(2)}_{4}+... .
\label{1.8}
\eea \ees
In the above $g_{1}$, $g_{3}$, etc. are functions of $x$ and $t$ and $g_{1}^{*}$, $g_{3}^{*}$, etc. are functions of $-x$ and $t$. Substituting the above series expansions in (4) and equating the coefficients of same powers of $\epsilon$ to zero we obtain a set of equations for the unknown functions $g(x,t)$, $g^{*}(-x,t)$, $f(x,t)$ and $f^{*}(-x,t)$.  By solving them recursively we can obtain the explicit form of these functions. Generally, when using this method, the expansion continues to infinite order in $\epsilon$. When we truncate the expansion at finite order one essentially obtains an approximate solution.  However, in the case of soliton possessing nonlinear evolution equations solvable by inverse scattering method, when performing the perturbation method for bilinear equations, an appropriate choice of $g_1$, $g_1^*$, etc. (which satisfy linear differential equations) makes the infinite expansion truncate with a finite number of terms, leading to an exact convergent soliton solution \cite{8,14,15}. We essentially use this method in our further analysis.
\section{One bright soliton solution}
The one soliton solution can be obtained by truncating the series expansion as $g(x,t)=\epsilon g_{1}+\epsilon^{3} g_{3}$, $g^{*}(-x,t)=\epsilon g_{1}^{*}+\epsilon^{3} g_{3}^{*}$, $f(x,t)=1+\epsilon^{2} f_{2}+\epsilon^{4} f_{4}$, $f^{*}(-x,t)=1+\epsilon^{2}f^{*}_{2}+\epsilon^{4}f^{*}_{4}$ with $s^{(1)}(-x,t)=\epsilon^{2}s^{(1)}_{2}$, $s^{(2)}(-x,t)=\epsilon^{2}s^{(2)}_{2}$, $h^{(1)}(-x,t)=\epsilon^{2}h^{(1)}_{2}$ and $h^{(2)}(-x,t)=\epsilon^{2}h^{(2)}_{2}$. Substituting these expansions in the bilinear equations and following the procedure given above we end up with a system of linear partial differential equations (PDEs) for the unknown functions $g_{1}$, $g_{3}$, $g^{*}_{1}$, $g^{*}_{3}$, $f_{2}$, $f_{4}$, $f^{*}_{2}$ and $f^{*}_{4}$. They are
\bes
\bea
&&\hspace{-0.5cm}D_1(g_1\cdot 1)=0,~D_2(g_1^{*}\cdot 1)=0,~D_3(1.f_2+f_2.1)=4h_2^{(1)}, \label{1.9a}\\
&&\hspace{-0.5cm}D_3(1.f^{*}_2+f^{*}_2.1)=4h_2^{(2)},~h_2^{(1)}=h_2^{(2)}=g_1\cdot g_1^{*}\label{1.9b}, \\
&&\hspace{-0.5cm}D_1(g_3\cdot 1+g_1\cdot f_2)=-2g^{*}_1s_2^{(1)}, ~D_2(g_3^{*}\cdot 1+g_1^{*}\cdot f^{*}_2)=2g_1s_2^{(2)}\label{1.9c},\\
&&\hspace{-0.5cm}D_3(1.f_4+f_4.1+f_2\cdot f_2)=4(h_2^{(1)}\cdot f_2+h_4^{(1)}),~s_2^{(1)}=g_1^2,\label{1.9d}\\
&&\hspace{-0.5cm}D_3(1.f^{*}_4+f^{*}_4.1+f^{*}_2\cdot f^{*}_2)=4(h_2^{(2)}\cdot f^{*}_2+h_4^{(2)}),~s_2^{(2)}=g_1^{*2}\\
&&\hspace{-0.5cm}h_4^{(1)}=-h_2^{(1)}\cdot f^{*}_2+g_1\cdot g_3^{*}+g_3g_1^{*},~~h_4^{(2)}=-h_2^{(2)}\cdot f_2+g_1\cdot g_3^{*}+g_3\cdot g_1^{*}.
\label{1.9e}
\eea \ees
The following form of $g_1(x,t)$, $g_1^{*}(-x,t)$, $f_2(x,t)$ and $f^{*}_2(-x,t)$ are consistent with Eqs. (\ref{1.9a}) -  (\ref{1.9b}), that is
\bes
\bea
&&g_1(x,t)=\al_1e^{\bar{\xi}_1},\quad g_1^{*}(-x,t)=\ba_1e^{\xi_1}, \\
&&f_2(x,t)=e^{\xi_1+\bar{\xi}_1+\del_1},\quad f^{*}_2(-x,t)=e^{\xi_1+\bar{\xi}_1+\del_1},
\eea \ees
where $\xi_1=i k_{1}x-ik_{1}^{2}t+\xi_1^{(0)}$, $\bar{\xi_1}=i \bar{k_{1}}x+i\bar{k_{1}^{2}}t+\bar{\xi}_1^{(0)}$, $e^{\del_1}=\frac{-2\al_{1}\ba_{1}}{\kappa}$ and $\kappa=(k_{1}+\bar{k_{1}})^2$. Here $k_1$, $\bar{k}_1$, $\al_1$, $\ba_1$, $\xi_1^{(0)}$ and $\bar{\xi}_1^{(0)}$ are arbitrary complex constants. The auxiliary functions are found to be
\bea
s^{(1)}_{2}=\al_{1}^{2}e^{2\bar{\xi_1}},\quad s^{(2)}_{2}=\ba_{1}^{2}e^{2\xi_1},\quad h^{(1)}_{2}=h^{(2)}_{2}=\al_{1}\ba_{1}e^{\xi_{1}+\bar{\xi_1}}.
\label{1.18}
\eea
 Substituting the above in Eqs. (\ref{1.9c}) - (\ref{1.9e}) and solving the resultant equations, one obtains 
\bea
&&\hspace{-1.0cm}g_3(x,t)=e^{\xi_1+2\bar{\xi_1}+\del_{11}},~g_3^{*}(-x,t)=e^{2\xi_1+\bar{\xi_1}+\Delta_{11}},~f_4=f^{*}_4=e^{2(\xi_1+\bar{\xi_1})+R}, 
\eea
where $e^{\del_{11}}=\frac{-\al_{1}^{2}\ba_{1}}{\kappa}$, $e^{\Del_{11}}=\frac{-\al_{1}\ba_{1}^{2}}{\kappa}$ and  $e^{R}=\frac{\al_{1}^{2}\beta_{1}^{2}}{\kappa^2}$. The auxiliary functions $h_4^{(1)}$ and $h_4^{(2)}$ are found to be zero. One may note that the auxiliary functions $h^{(1)}_{2}$ and $h^{(2)}_{2}$ do not differ from each other. Our results also show that the functions $f(x,t)$ and $f^{*}(-x,t)$ are one and the same at all orders, that is $f(x,t)=f^{*}(-x,t)$. Consequently, three auxiliary functions are sufficient to  bilinearize Eq. (\ref{1.1}).
Substituting the above expressions back in the relevant expressions given in (6) and rewriting them suitably we arrive at the following general one soliton solution of (\ref{1.1}), 
%$s^{(1)}_{2}$, $s^{(2)}_{2}$, $h^{(1)}_{2}$ and $h^{(2)}_{2}$
\bea
q(x,t)=\frac{\al_1e^{\bar{\xi}_1}+e^{\xi_1+2\bar{\xi_1}+\del_{11}}}{1+e^{\xi_1+\bar{\xi_1}+\del_1}+e^{2(\xi_1+\bar{\xi_1})+R}}\equiv \frac{\al_1e^{\bar{\xi}_1}}{1+e^{\xi_1+\bar{\xi_1}+\Del}},~e^{\Del}=\frac{-\al_1\ba_1}{\kappa}.
\label{1.15}
\eea
From the bilinear form of parity transformed complex conjugate equation we can obtain the parity ($\cal{P}$) transformed complex conjugate field in the form 
\bea
q^{*}(-x,t)=\frac{\ba_1e^{\xi_1}+e^{2\xi_1+\bar{\xi_1}+\Delta_{11}}}{1+e^{\xi_1+\bar{\xi_1}+\del_1}+e^{2(\xi_1+\bar{\xi_1})+R}}\equiv\frac{\ba_1e^{\xi_1}}{1+e^{\xi_1+\bar{\xi_1}+\Del}}.
\label{1.16}
\eea 
\begin{figure}[h]
\centering
\includegraphics[width=0.35\linewidth]{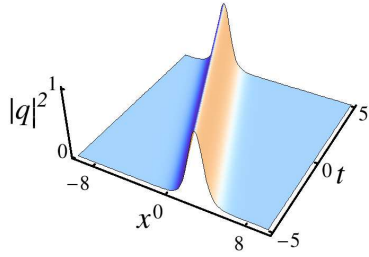}
\caption{Non-singular most general one bright soliton solution of Eq. ({\ref{1.1}})}
\label{fig1}
\end{figure}
Indeed one can check the correctness of the solutions (\ref{1.15}) and (\ref{1.16}) by substituting them back in Eqs. (\ref{1.1}) and (\ref{1.2}). We note here that the small parameter $\epsilon$  has been absorbed into the arbitrary constants $\xi_1^{(0)}$ and $\bar{\xi}_1^{(0)}$ by rewriting them as $\xi_1^{(0)}+\ln\epsilon$ and $\bar{\xi}_1^{(0)}+\ln\epsilon$. This is equivalent to effectively choosing the parameter as one at the end of the calculations. The above one bright soliton solution (\ref{1.15}) can also be  rewritten as 
\bea
q(x,t)=\frac{\al_1e^{-\frac{\Del}{2}}e^{\frac{(\bar{\xi}_{1R}-\xi_{1R})}{2}+i\frac{(\bar{\xi}_{1I}-\xi_{1I})}{2}}}{2[\cosh(\chi_1)\cos(\chi_2)+i\sinh(\chi_1)\sin(\chi_2)]},
\eea
and a similar expression for the parity transformed complex conjugate field is,
\bea
q^*(-x,t)=\frac{\ba_1e^{-\frac{\Del}{2}}e^{-\frac{(\bar{\xi}_{1R}-\xi_{1R})}{2}-i\frac{(\bar{\xi}_{1I}-\xi_{1I})}{2}}}{2[\cosh(\chi_1)\cos(\chi_2)+i\sinh(\chi_1)\sin(\chi_2)]},
\eea
where $\chi_1= \frac{\bar{\xi}_{1R}+\xi_{1R}+\Del_R}{2}$, $\chi_2=\frac{\bar{\xi}_{1I}+\xi_{1I}+\Del_I}{2}$, $\xi_{1I}=k_{1R}x+(k_{1I}^{2}-k_{1R}^{2})t$, $\bar{\xi}_{1I}=\bar{k}_{1R}x+(\bar{k}_{1R}^{2}-\bar{k}_{1I}^{2})t$, $\xi_{1R}=-k_{1I}(x-2k_{1R}t)$, $\bar{\xi}_{1R}=\bar{k}_{1I}(x+2\bar{k}_{1R}t)$, $\Del_R=\frac{1}{2}\log{\bigg(\frac{|\al_1|^2|\ba_1|^2}{|k_1+\bar{k}_1|^2}\bigg)}$ and $\Del_I=\frac{-i}{2}\log\bigg(\frac{\al_1\ba_1(k_1^{*}+\bar{k}_1^*)^2}{\al_1^*\ba_1^*(k_1+\bar{k}_1)^2}\bigg)$. To our knowledge the solution given above is more general than the one reported in the literature so far and which is in general non-singular. As a typical example, we display in Figure 1 the non-singular one soliton solution for the wave  parameters $k_1=0.4+i$, $\bar{k}_1=-0.4+i$, $\al_1=1+i$ and $\ba_1=1-i$. The special feature of the obtained one bright soliton solution is that it is invariant under all the symmetry operations which are given in Ref. \cite{4}. We also note here that, however, for the choice of specific parametric condition
\bea
2(k_{1R}k_{1I}+\bar{k}_{1R}\bar{k}_{1I})((2n+1)\pi-\Del_I)=-\Del_R(k_{1I}^2-k_{1R}^2+\bar{k}_{1R}^2-\bar{k}_{1I}^2),
\eea
 where $n = 0,1,2,...$, the soliton becomes singular for finite value of $t$ at $x = 0$.
 Unlike the standard NLS equation, whose one soliton solution contains only two complex free parameters, the solution (\ref{1.15}) given above is characterized by four complex parameters $k_1$, $\bar{k_1}$, $\al_1$ and $\ba_1$. The central position and speed of the soliton are given by $\frac{\Del_R}{2(\bar{k}_{1I}-k_{1I})}$ and $2k_{1R}$, respectively. Different kinds of localized solutions have also been  reported in the literature such as periodic and hyperbolic soliton solutions \cite{9}, dark and anti-dark soliton solutions \cite{10}, breather solutions \cite{11} and rational soliton solutions \cite{12}   

 The two parameter family of breathing one soliton solution which is reported in \cite{4} can be extracted from (\ref{1.15}) by restricting the wave numbers $k_{1}$ and $\bar{k}_{1}$ to be pure imaginary, that is $k_{1}=i2\eta_1$ and $\bar{k}_{1}=i2\bar{\eta}_1$ and considering $\al_{1}=-2(\eta_1+\bar{\eta}_1)e^{i\bar{\theta}_{1}}$ and $\ba_{1}=-2(\eta_1+\bar{\eta}_1)e^{i\theta_{1}}$ (where $\eta_1$, $\bar{\eta}_1$, $\theta_1$ and $\bar{\theta}_1$, are all real) in it. The resultant action yields
\begin{eqnarray}
q(x,t)=-\frac{2(\eta_{1}+\bar{\eta_1})e^{i\bar{\theta}_1}e^{-4i\bar{\eta_{1}}^{2}t}e^{-2\bar{\eta_1}x}}{1+e^{i(\theta_1+\bar{\theta}_1)}e^{4i(\eta_{1}^{2}-\bar{\eta_{1}}^{2})t}e^{-2(\eta_{1}+\bar{\eta_1})x}},
\label{b}
\end{eqnarray}
 as obtained in \cite{4}. Imposing the same restrictions on (\ref{1.16}) we can obtain the expression for the parity transformed complex conjugate field whose expression also matches with the one given in \cite{4}, that is
\begin{eqnarray}
q^*(-x,t)=-\frac{2(\eta_{1}+\bar{\eta_1})e^{i\theta_1}e^{4i\eta_1^{2}t}e^{-2\eta_1x}}{1+e^{i(\theta_1+\bar{\theta}_1)}e^{4i(\eta_{1}^{2}-\bar{\eta_{1}}^{2})t}e^{-2(\eta_{1}+\bar{\eta_1})x}}.
\end{eqnarray} 
As reported in Ref. \cite{4}, the two parameter breathing soliton solution (\ref{b}) develops singularity in a finite time as may be checked from condition (13) for the choice $k_{1R} = \bar{k}_{1R} = 0$, $k_{1I}=2\eta_1$, $\bar{k}_{1I}=2\bar{\eta}_1$, $\al_{1}=-2(\eta_1+\bar{\eta}_1)e^{i\bar{\theta}_{1}}$ and $\ba_{1}=-2(\eta_1+\bar{\eta}_1)e^{i\theta_{1}}$. We can capture the standard envelope soliton solution of NLS equation, that is
\begin{eqnarray}
q(x,t)=-2\eta_1e^{-i(\theta_1+4\eta_1^{2}t)}\sech(2\eta_1x).
\end{eqnarray}
by substituting $\eta_1=\bar{\eta}_1$ and $\theta_1=-\bar{\theta}_1$ in (\ref{1.15}) with the above mentioned restrictions. From the above analysis, it is clear that the suggested procedure is valid for the general case, that is $q^*(-x,t)\neq q(x,t)$.
\section{Two bright soliton solution}
The two soliton solution can be obtained by truncating the series expansion, $g(x,t)=\epsilon g_{1}+\epsilon^{3} g_{3}+\epsilon^{5} g_{5}+\epsilon^{7} g_{7}$, $g^{*}(-x,t)=\epsilon g_{1}^{*}+\epsilon^{3} g_{3}^{*}+\epsilon^{5} g_{5}^{*}+\epsilon^{7} g_{7}^{*}$, $f(x,t)=1+\epsilon^{2} f_{2}+\epsilon^{4} f_{4}+\epsilon^{6} f_{6}+\epsilon^{8} f_{8}$, $f^{*}(-x,t)=1+\epsilon^{2}f^{*}_{2}+\epsilon^{4} f^{*}_{4}+\epsilon^{6} f^{*}_{6}+\epsilon^{8}f^{*}_{8}$ and the auxiliary functions can be expanded in the series expansion as $s^{(1)}(-x,t)=\epsilon^{2}s^{(1)}_{2}+\epsilon^{4}s^{(1)}_{4}+\epsilon^{6}s^{(1)}_{6}$, $s^{(2)}(-x,t)=\epsilon^{2}s^{(2)}_{2}+\epsilon^{4}s^{(2)}_{4}+\epsilon^{6}s^{(2)}_{6}$, $h^{(1)}(-x,t)=\epsilon^{2}h^{(1)}_{2}+\epsilon^{4}h^{(1)}_{4}+\epsilon^{6}h^{(1)}_{6}$, $h^{(2)}(-x,t)=\epsilon^{2}h^{(2)}_{2}+\epsilon^{4}h^{(2)}_{4}+\epsilon^{6}h^{(2)}_{6}$. Substituting these expansions in the bilinear Eqs. (\ref{1.5})-(\ref{1.6}), we get a system of PDEs for the functions that appear in the series expansion. Solving the resultant equations consistently we can obtain the explicit forms of the functions $g(x,t)$, $g^{*}(-x,t)$ and $f(x,t)$ in the following form,
\bes
\bea
g(x,t)&=&\al_{1}e^{\bar{\xi_1}}+\al_{2}e^{\bar{\xi_2}}+e^{\xi_{1}+2\bar{\xi_1}+\Del_1}+e^{\xi_{2}+2\bar{\xi_1}+\Del_2}+e^{\xi_{1}+2\bar{\xi_2}+\Del_3}+e^{\xi_{2}+2\bar{\xi_2}+\Del_4} \nonumber \\&&
+e^{\xi_{1}+\bar{\xi_1}+\bar{\xi_2}+\Del_{11}}+e^{\xi_{2}+\bar{\xi_1}+\bar{\xi_2}+\Del_{12}}+e^{2\xi_{1}+2\bar{\xi_1}+\bar{\xi_2}+\Del_{21}}+e^{2\xi_{1}+\bar{\xi_1}+2\bar{\xi_2}+\Del_{22}}\nonumber\\&&
+e^{2\xi_{2}+2\bar{\xi_1}+\bar{\xi_2}+\Del_{23}}+e^{2\xi_{2}+\bar{\xi_1}+2\bar{\xi_2}+\Del_{24}}+e^{\xi_{1}+\bar{\xi_1}+\xi_2+2\bar{\xi_2}+\Del_{25}}\nonumber\\&&
+e^{\xi_{1}+2\bar{\xi_1}+\xi_2+\bar{\xi_2}+\Del_{26}}+e^{2\xi_{1}+2\bar{\xi_1}+\xi_{2}+2\bar{\xi_2}+\Del_{31}}+e^{2\bar{\xi_1}+2\bar{\xi_2}+\xi_{1}+2\xi_{2}+\Del_{32}}
\label{3.1}
\eea
\bea
f(x,t)&=&1+e^{\xi_{1}+\bar{\xi_1}+\del_1}+e^{\xi_{2}+\bar{\xi_1}+\del_2}+e^{\xi_{1}+\bar{\xi_2}+\del_3}+e^{\xi_{2}+\bar{\xi_2}+\del_4}+e^{2(\xi_{1}+\bar{\xi_1})+\del_{11}}\nonumber\\&&
+e^{2(\xi_{2}+\bar{\xi_1})+\del_{12}}+e^{2(\xi_{1}+\bar{\xi_2})+\del_{13}}+e^{2(\xi_{2}+\bar{\xi_2})+\del_{14}}+e^{2\bar{\xi_1}+\xi_{1}+\xi_{2}+\del_{15}}\nonumber\\&&
+e^{2\bar{\xi_2}+\xi_{1}+\xi_{2}+\del_{16}}+e^{2\xi_{1}+\bar{\xi_1}+\bar{\xi_2}+\del_{17}}+e^{2\xi_{2}+\bar{\xi_1}+\bar{\xi_2}+\del_{18}}+e^{\xi_{1}+\bar{\xi_1}+\xi_{2}+\bar{\xi_2}+\del_{19}}\nonumber\\&&
+e^{2\xi_{1}+2\bar{\xi_1}+\xi_{2}+\bar{\xi_2}+\del_{21}}+e^{2\xi_{1}+\bar{\xi_1}+\xi_{2}+2\bar{\xi_2}+\del_{22}}+e^{\xi_{1}+2\bar{\xi_1}+2\xi_{2}+\bar{\xi_2}+\del_{23}}\nonumber\\&&
+e^{\xi_{1}+\bar{\xi_1}+2\xi_{2}+2\bar{\xi_2}+\del_{24}}+e^{2(\xi_{1}+\bar{\xi_1}+\xi_{2}+\bar{\xi_2})+\del_{31}}\equiv f^{*}(-x,t),
\label{3.2}\eea
\bea
g^{*}(-x,t)&=&\ba_{1}e^{\xi_{1}}+\ba_{2}e^{\xi_{2}}+e^{2\xi_{1}+\bar{\xi_1}+\ga_1}+e^{2\xi_{1}+\bar{\xi_2}+\ga_2}+e^{2\xi_{2}+\bar{\xi_1}+\ga_3}+e^{2\xi_{2}+\bar{\xi_2}+\ga_4} \nonumber \\&&
+e^{\xi_{1}+\bar{\xi_1}+\xi_{2}+\ga_{11}}+e^{\xi_{1}+\xi_{2}+\bar{\xi_2}+\ga_{12}}+e^{2\xi_{1}+2\bar{\xi_1}+\xi_{2}+\ga_{21}}+e^{\xi_{1}+2\bar{\xi_1}+2\xi_{2}+\ga_{22}}\nonumber\\&&
+e^{2\xi_{1}+2\bar{\xi_2}+\xi_{2}+\ga_{23}}+e^{2\xi_{2}+2\bar{\xi_2}+\xi_{1}+\ga_{24}}+e^{2\xi_{1}+\bar{\xi_1}+\xi_{2}+\bar{\xi_2}+\ga_{25}}\nonumber\\&&
+e^{\xi_{1}+\bar{\xi_1}+2\xi_{2}+\bar{\xi_2}+\ga_{26}}+e^{2\xi_{1}+2\bar{\xi_1}+2\xi_{2}+\bar{\xi_2}+\ga_{31}}+e^{2\xi_{1}+\bar{\xi_1}+2\xi_{2}+2\bar{\xi_2}+\ga_{32}}.
\eea\label{3.3} \ees
Substituting the expressions (19) in (\ref{1.4}) we obtain the two soliton solution of (\ref{1.1}) and the  parity ($\cal{P}$) transformed complex conjugate field  given in (\ref{1.4})
\begin{figure}[ht]
\centering
\includegraphics[width=0.35\linewidth]{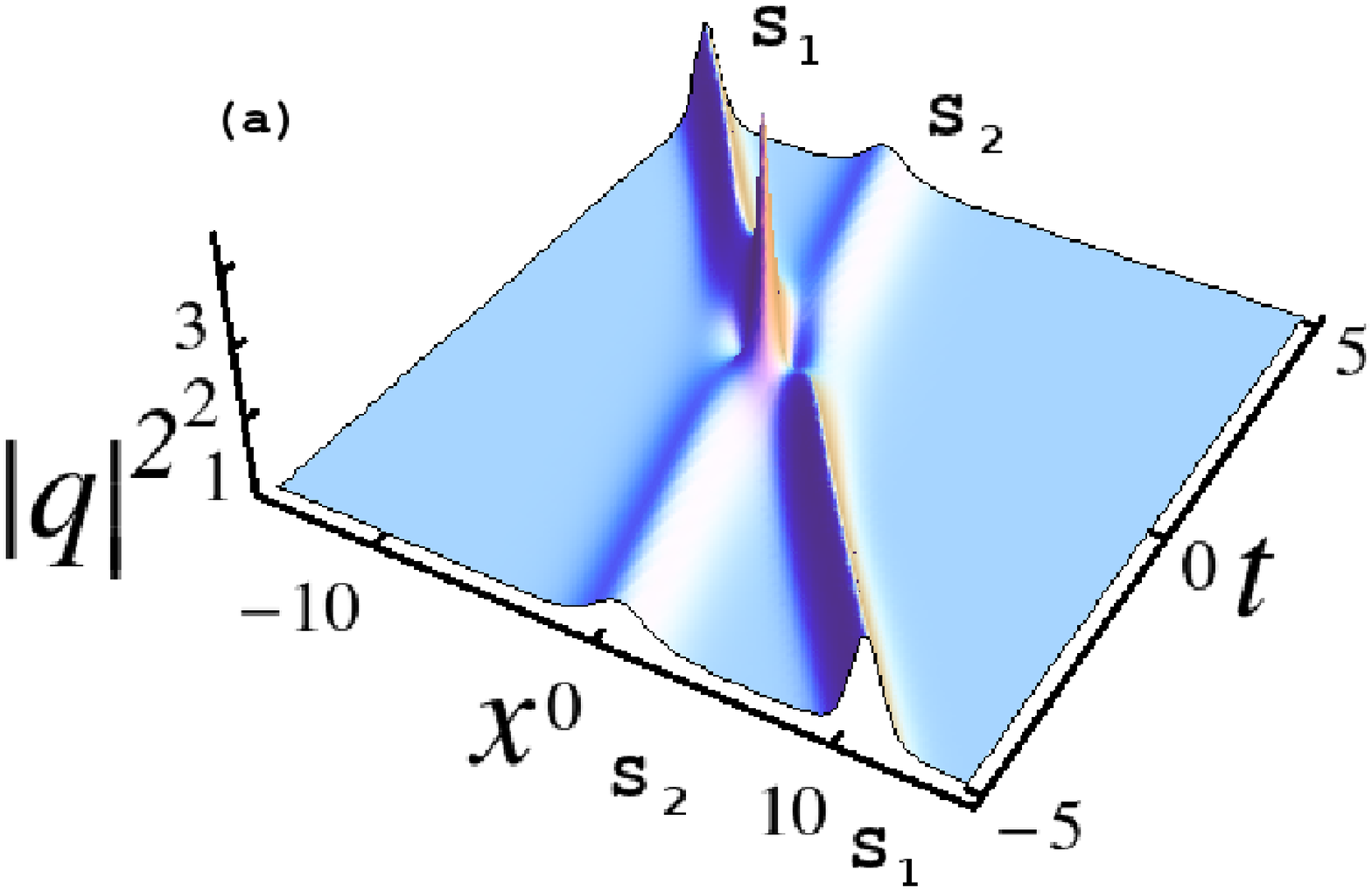}~~\includegraphics[width=0.35\linewidth]{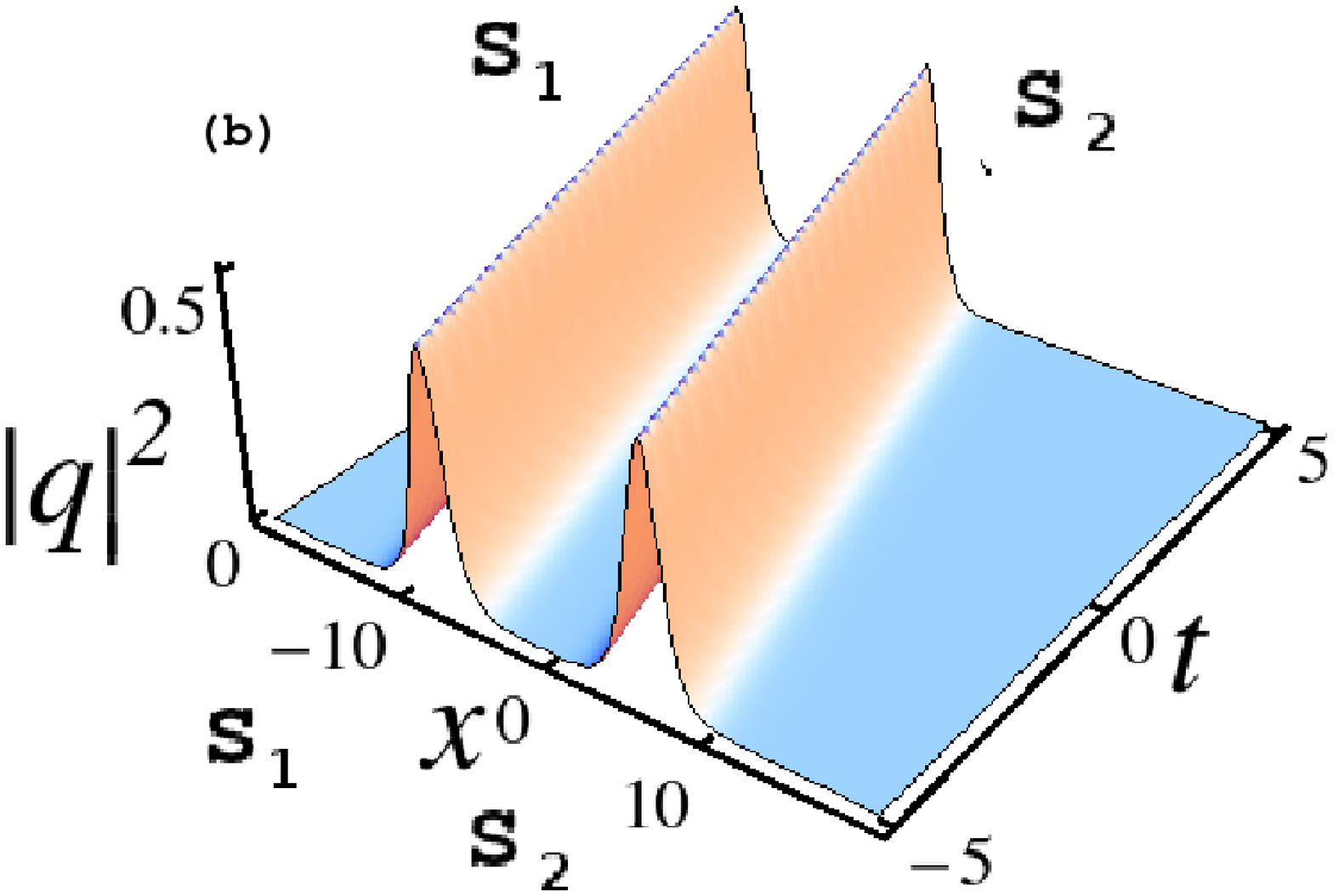}\\
\includegraphics[width=0.35\linewidth]{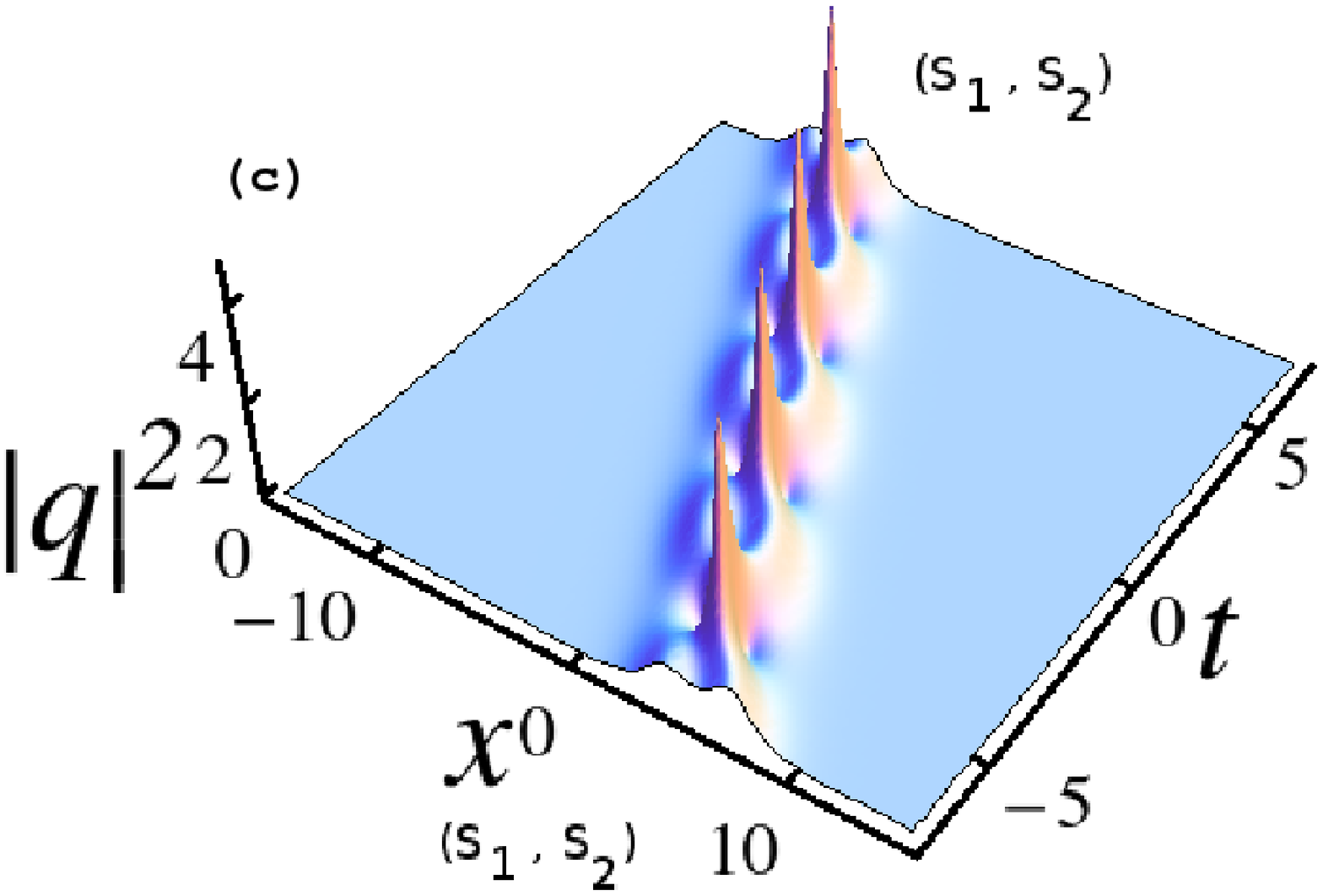}~~\includegraphics[width=0.35\linewidth]{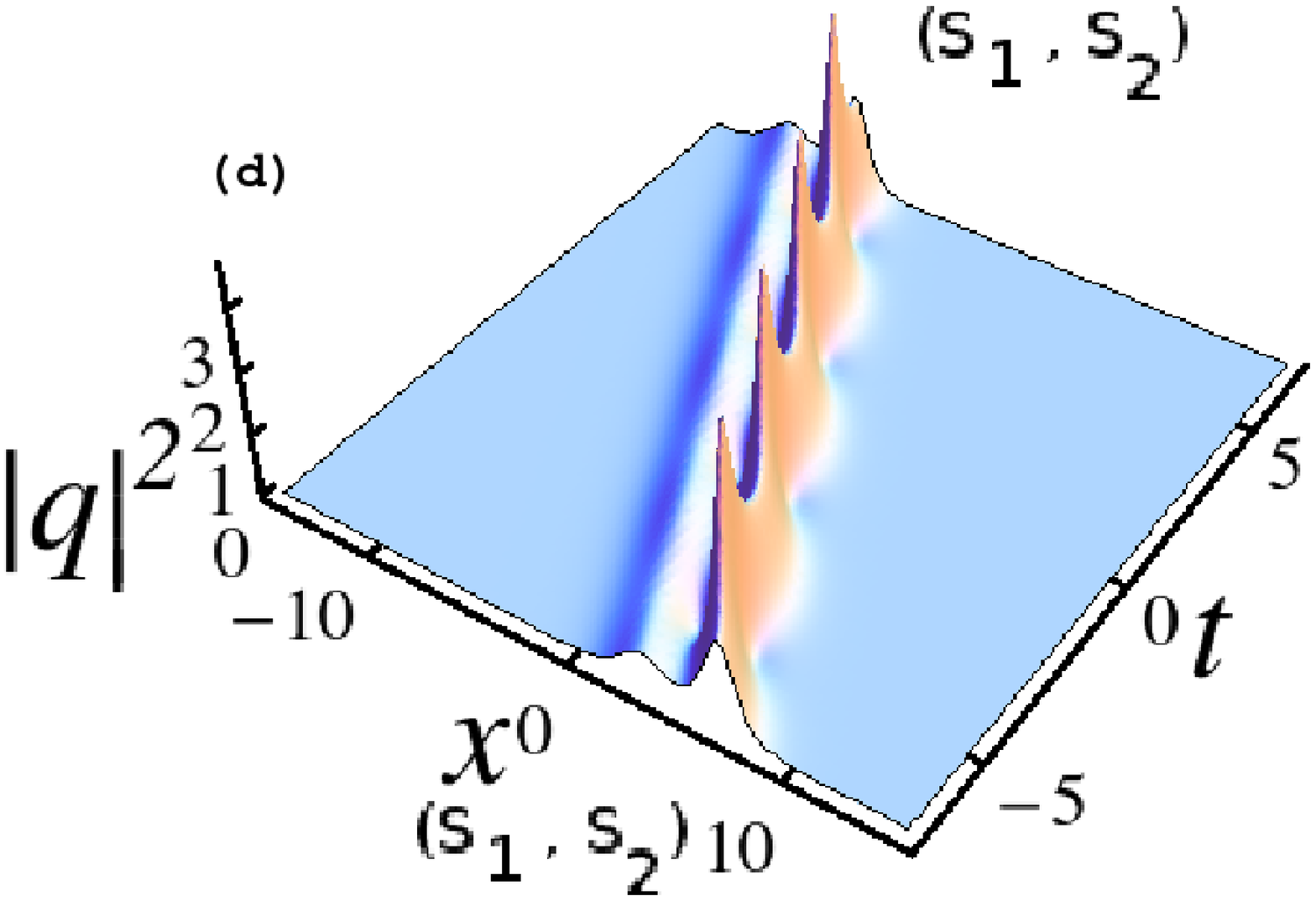}
\caption{(a) Elastic collision between two nonlocal solitons. (b) Parallel propagation of two solitonic bound state interaction. (c) Breathing type bound state. (d) Suppression of oscillation in the breathing type bound state}
\label{fig2}
\end{figure}
with $\xi_{j}=i k_{j}x-ik_{j}^{2}t+\xi_j^{(0)}$, $\bar{\xi_j}=i \bar{k_{j}}x+i\bar{k_{j}^{2}}t+\bar{\xi}_j^{(0)}$, $j=1,2$, which can again be checked by substituting back in (\ref{1.1}) and (\ref{1.2}). The auxiliary functions and the explicit expressions of all the parameters that appear in (19) are given in the Appendix. The two soliton solution is characterized by eight complex parameters, namely $k_1$, $\bar{k_1}$, $k_2$, $\bar{k_2}$, $\al_1$, $\al_2$, $\ba_1$ and $\ba_2$. One can also check that the two soliton solution obtained by Ablowitz and Musslimani \cite{4} through the inverse scattering formalism can be obtained for the case that the real parts of the parameters $k_{1R}$, $\bar{k}_{1R}$, $k_{2R}$ and $\bar{k}_{2R}$ vanish and that the imaginary parts take the values $k_{1I}=2\eta_1$, $\bar{k}_{1I}=2\bar{\eta}_1$, $k_{2I}=2\eta_2$ and $\bar{k}_{2I}=2\bar{\eta}_2$ in the above. 

 The various types of interactions which can occur between the associated two solitons can be characterized through these eight parameters. In Figure 2(a) we display the head-on collision between the two nonlocal solitons. We fix the parameters as $k_{1}=-0.4+0.7i$, $\bar{k}_1=0.4+0.7i$, $k_{2}=-1.2+1.2i$, $\bar{k}_{2}=-1.2+1.2i$, $\al_{1}=0.5+0.5i$, $\ba_{1}=0.5-0.5i$, $\al_{2}=0.5+0.5i$ and $\ba_{2}=0.5-0.5i$. The velocities are given by the real parts of the wave numbers, namely $k_{1}$, $\bar{k}_{1}$, $k_{2}$ and $\bar{k}_{2}$. As expected, the larger amplitude soliton is moving faster than the smaller amplitude soliton. At a large negative value of $t$, the two solitons are well separated and they are localized around $x=2k_{1R}t$ and $x=2k_{2R}t$, respectively. Around $x=0$, the two solitons interact with each other and after a finite time these two solitons separate from each other and travel with their own velocities ($2k_{1R}$ and $2k_{2R}$), respectively. After the interaction, the velocities of either of the solitons is not altered which in turn confirms the elastic nature of interaction that occurred between these two nonlocal solitons. The effect of interaction is reflected  only in their phases and hence the  energy is conserved during the evolution process.  

 Next we investigate different types of bound states in the NNLS equation as observed in the case of coherently coupled nonlinear Schr\"{o}dinger system in Ref. \cite{13}. As we pointed out earlier the velocities of the solitons are determined by the parameters $2k_{1R}$ and $2k_{2R}$ and the central positions of the solitons are given by the parameters $\frac{\Del_R}{2(\bar{k}_{1I}-k_{1I})}$ and $\frac{\Del_R}{2(\bar{k}_{2I}-k_{2I})}$, respectively. To explore the two soliton bound state in the NNLS equation, we fix the soliton parameters $k_{1R}$ and $k_{2R}$ as equal so that the two nonlocal solitons can propagate with the same velocity. To obtain a parallel soliton propagation, we fix the values of $k_{1I}$ and $k_{2I}$ to be nearly equal, that is $k_{1I}=0.7$ and $k_{2I}=0.71$ and the other parameters as $k_{1R}=-0.5$, $\bar{k}_{1R}=0.5$, $k_{2R}=-0.5$, $\bar{k}_{2R}=0.5$, $\bar{k}_{1I}=0.7$, $\bar{k}_{2I}=0.71$,  $\al_{1}=1+i$, $\ba_{1}=1-i$, $\al_{2}=1+i$ and $\ba_{2}=1-i$. By restricting the parameters in the above manner we can check the solitons propagating parallel to each other as seen in Figure 2(b). To explore the breathing type of bound state in the NNLS equation we fix the parameters as $k_{1}=-0.5+0.7i$, $\bar{k}_1=0.5+0.7i$, $k_2=0.5+1.5i$, $\bar{k}_{2}=-0.5+1.5i$, $\al_{1}=1+i$, $\ba_{1}=1-i$, $\al_{2}=1+i$ and $\ba_{2}=1-i$. The outcome is demonstrated in Figure 2(c). As one can see even though the central positions of the two solitons are different, they are bound together and travel like a single solitonic state. We can control the oscillatory behavior of these two solitons by tuning the complex parameters $\al_{1}$, $\ba_{1}$, $\al_{2}$ and $\ba_{2}$. In Figure 2(c), we demonstrate the oscillations that occur in the amplitude of the second soliton. One can also induce the oscillations in the first soliton by tuning the above parameters. The suppression of oscillations in the first soliton is demonstrated in Figure 2(d). To obtain this, we fix the parameters as $k_{1}=-0.5+0.7i$, $\bar{k}_1=0.5+0.7i$, $k_2=0.5+1.5i$, $\bar{k}_{2}=-0.5+1.5i$, $\al_{1}=0.5+0.5i$, $\ba_{1}=0.5-0.5i$, $\al_{2}=1+i$ and $\ba_{2}=1-i$. If we slightly increase the real part of the second wave number, that is $k_{2R}$ and $\bar{k}_{2R}$, we can recover the colliding solitons. We can also induce the oscillations in the second soliton by tuning the parameters $\al_2$ and $\ba_2$.       

\section{Conclusion}

In this work, we have thus developed a nonstandard bilinearization to construct more general  $\cal{PT}$-invariant soliton solutions for the NNLS Eq. (\ref{1.1}). To achieve this, we have considered a novel non-standard bilinearization procedure in which we treated the nonlinear Schr\"{o}dinger field and its associated parity transformed complex conjugate field as separate entities. The obtained soliton solutions are invariant under the symmetry operations given in Ref. \cite{4}. One can deduce $N$-soliton solutions for the NNLS equation by proceeding with higher order functions. Using this procedure, one can also deduce the dark soliton solutions for the defocusing case of NNLS and one can consider bilinearization of coupled NNLS equations as well. These results will be presented elsewhere.  
\section*{Acknowledgements}
The work of SS supported by the Department of Atomic Energy (DAE), Government of India, under the Grant No. 10/9/2010/RRF-R\&D-II/8756. The work of MS forms part of a research project sponsored by DST-SERB, Government of India, under the Grant No. EMR/2016/001818. The research work of ML is supported by a NASI Senior Scientist Platinum Jubilee Fellowship (NAS 69/5/2016-17) and forms part of the DAE-NBHM research project (2/48 (5)/2015/NBHM (R.P.)/R\&D-II/14127).

\section*{Appendix}
In the following, we define the various quantities that appear in the two soliton solutions (19) of the NNLS: 
\bea
&&\nonumber e^{\Del_1}=-\al_{1}^{2}\ba_{1}\kappa_{11},~e^{\Del_2}=-\al_{1}^{2}\ba_{2}\kappa_{21},~e^{\Del_3}=-\al_{2}^{2}\ba_{1}\kappa_{12},~e^{\Del_4}=-\al_{2}^{2}\ba_{2}\kappa_{22},\\
&&\nonumber e^{\Del_{11}}=-2\al_{1}\al_{2}\ba_{1}(k_{1}^{2}+\bar{k_{1}}^{2}+\bar{k_{2}}^{2}+k_{1}\bar{k_1}+k_{1}\bar{k_2}-\bar{k_1}\bar{k_2})\kappa_{11}\kappa_{12},\\
&&\nonumber e^{\Del_{12}}=-2\al_{1}\al_{2}\ba_{2}(k_{2}^{2}+\bar{k_{1}}^{2}+\bar{k_{2}}^{2}+k_{2}\bar{k_1}+k_{2}\bar{k_2}-\bar{k_1}\bar{k_2})\kappa_{21}\kappa_{22},\\
&&\nonumber e^{\Del_{21}}=\al_{1}^{2}\al_{2}\ba_{1}^{2}\bar{\varrho}_{12}\kappa_{11}^2\kappa_{12},~e^{\Del_{22}}=\al_{1}\al_{2}^{2}\ba_{1}^{2}\bar{\varrho}_{12}\kappa_{11}\kappa_{12}^2,~e^{\Del_{23}}=\al_{1}^{2}\al_{2}\ba_{2}^{2}\bar{\varrho}_{12}\kappa_{21}^2\kappa_{22},\\
&&\nonumber e^{\Del_{24}}=\al_{1}\al_{2}^{2}\ba_{2}^{2}\bar{\varrho}_{12}\kappa_{21}\kappa_{22}^2,~e^{\Del_{31}}=-\al_{1}^{2}\al_{2}^{2}\ba_{1}^{2}\ba_{2}\varrho_{12}\bar{\varrho}_{12}^2\kappa_{11}^2\kappa_{21}\kappa_{12}^2\kappa_{22},\\
&&\nonumber e^{\Del_{25}}=2\al_{1}\al_{2}^{2}\ba_{1}\ba_{2}(k_{1}^{2}+\bar{k_1}^{2}+k_{2}^{2}+k_{1}(\bar{k_{1}}-k_{2})+k_{2}\bar{k_1})\bar{\varrho}_{12}\kappa_{11}\kappa_{21}\kappa_{12}\kappa_{22},\\
&&\nonumber e^{\Del_{26}}=2\al_{1}^2\al_{2}\ba_{1}\ba_{2}(k_{1}^{2}+\bar{k_2}^{2}+k_{2}^{2}+k_{1}(\bar{k_{2}}-k_{2})+k_{2}\bar{k_2})\bar{\varrho}_{12}\kappa_{11}\kappa_{21}\kappa_{12}\kappa_{22},\\
&&\nonumber e^{\Del_{32}}=-\al_{1}^{2}\al_{2}^{2}\ba_{1}\ba_{2}^{2}\varrho_{12}\bar{\varrho}_{12}^2\kappa_{11}\kappa_{21}^2\kappa_{12}\kappa_{22}^2,~e^{\del_1}=-2\al_{1}\ba_{1}\kappa_{11},~e^{\del_2}=-2\al_{1}\ba_{2}\kappa_{21},\\
&&\nonumber e^{\del_3}=-2\al_{2}\ba_{1}\kappa_{12},~e^{\del_4}=-2\al_{2}\ba_{2}\kappa_{22},~e^{\del_{11}}=\al_{1}^{2}\ba_{1}^{2}\kappa_{11}^2,~e^{\del_{12}}=\al_{1}^{2}\ba_{2}^{2}\kappa_{21}^2,\\
&&\nonumber e^{\del_{13}}=\al_{2}^{2}\ba_{1}^{2}\kappa_{12}^2,~e^{\del_{14}}=\al_{2}^{2}\ba_{2}^{2}\kappa_{22}^2,~e^{\del_{15}}=2\al_{1}^{2}\ba_{1}\ba_{2}\kappa_{11}\kappa_{21},~e^{\del_{16}}=2\al_{2}^{2}\ba_{1}\ba_{2}\kappa_{12}\kappa_{22},\\
&&\nonumber e^{\del_{17}}=2\al_{1}\al_{2}\ba_{1}^{2}\kappa_{11}\kappa_{12},~e^{\del_{18}}=2\al_{1}\al_{2}\ba_{2}^{2}\kappa_{21}\kappa_{22},~e^{\del_{19}}=4\al_{1}\al_{2}\ba_{1}\ba_{2}\kappa_{11}\kappa_{21}\kappa_{12}\kappa_{22}\nu_1,\\
&&\nu_1=\{\bar{k_1}^{2}(k_{2}^{2}+k_{2}\bar{k_{2}}+\bar{k_2}^{2})+k_{2}^{2}\bar{k_2}^{2}+\bar{k_1}k_{2}\bar{k_2}(\bar{k_2}-k_{2})+k_{1}^{2}(\bar{k_1}^{2}+k_{2}^{2}+\bar{k_2}^{2}\nonumber \\&& \hspace{1.0cm}
+\bar{k_2}k_{2}+\bar{k_1}(k_{2}-\bar{k_2}))+k_{1}(k_{2}\bar{k_2}(k_{2}-\bar{k_{2}})+\bar{k_1}^{2}(\bar{k_2}-k_{2})+\bar{k_1}(k_{2}^{2}\nonumber \\&& \hspace{1.0cm}
+6k_{2}\bar{k_2}+\bar{k_2}^{2}))\}\nonumber ,\\
&&\nonumber e^{\del_{21}}=-2\al_{1}^{2}\al_{2}\ba_{1}^{2}\ba_{2}\varrho_{12}\bar{\varrho}_{12}\kappa_{11}^2\kappa_{12}\kappa_{21}\kappa_{22},~e^{\del_{22}}=-2\al_{1}\al_{2}^{2}\ba_{1}^{2}\ba_{2}\varrho_{12}\bar{\varrho}_{12}\kappa_{11}\kappa_{12}^2\kappa_{21}\kappa_{22},\\
&&\nonumber e^{\del_{23}}=-2\al_{1}^{2}\al_{2}\ba_{1}\ba_{2}^{2}\varrho_{12}\bar{\varrho}_{12}\kappa_{11}\kappa_{12}\kappa_{21}^2\kappa_{22},~e^{\del_{24}}=-2\al_{1}\al_{2}^{2}\ba_{1}\ba_{2}^{2}\varrho_{12}\bar{\varrho}_{12}\kappa_{11}\kappa_{12}\kappa_{21}\kappa_{22}^2,~\\
&&\nonumber e^{\del_{31}}=\al_{1}^{2}\al_{2}^{2}\ba_{1}^{2}\ba_{2}^{2}\varrho_{12}^2\bar{\varrho}_{12}^2\kappa_{11}^2\kappa_{12}^2\kappa_{21}^2\kappa_{22}^2,~e^{\ga_1}=-\al_{1}\ba_{1}^{2}\kappa_{11},~ e^{\ga_2}=-\al_{2}\ba_{1}^{2}\kappa_{12},\\
&&\nonumber e^{\ga_3}=-\al_{1}\ba_{2}^{2}\kappa_{21},~e^{\ga_4}=-\al_{2}\ba_{2}^{2}\kappa_{22},~e^{\ga_{21}}=\al_{1}^{2}\ba_{1}^{2}\ba_{2}\varrho_{12}\kappa_{11}^2\kappa_{21},\\
&&\nonumber e^{\ga_{11}}=-2\al_{1}\ba_{1}\ba_{2}(k_{1}^{2}+k_{2}^{2}+\bar{k_1}^{2}+k_{1}(\bar{k_1}-k_{2})+k_{2}\bar{k_1})\kappa_{11}\kappa_{21},\\
&&\nonumber e^{\ga_{12}}=-2\al_{2}\ba_{1}\ba_{2}(k_{1}^{2}+k_{2}^{2}+\bar{k_2}^{2}+\bar{k_2}(k_{1}+k_{2})-k_{2}k_{1})\kappa_{12}\kappa_{22},\\
&&\nonumber e^{\ga_{22}}=\al_{1}^{2}\ba_{1}\ba_{2}^{2}\varrho_{12}\kappa_{11}\kappa_{21}^2,~e^{\ga_{23}}=\al_{2}^{2}\ba_{1}^{2}\ba_{2}\varrho_{12}\kappa_{12}^2\kappa_{22},~e^{\ga_{24}}=\al_{2}^{2}\ba_{1}\ba_{2}^{2}\varrho_{12}\kappa_{12}\kappa_{22}^2,\\
&&\nonumber e^{\ga_{25}}=2\al_{1}\al_{2}\ba_{1}^{2}\ba_{2}\varrho_{12}(k_{2}^{2}+\bar{k_1}^{2}+\bar{k_2}^{2}-\bar{k_1}\bar{k_2}+k_{2}(\bar{k_2}+\bar{k_1}))\kappa_{11}\kappa_{21}\kappa_{12}\kappa_{22},\\
&&\nonumber e^{\ga_{26}}=2\al_{1}\al_{2}\ba_{1}\ba_{2}^{2}\varrho_{12}(k_{1}^{2}+\bar{k_1}^{2}+\bar{k_2}^{2}-\bar{k_1}\bar{k_2}+k_{1}(\bar{k_2}+\bar{k_1}))\kappa_{11}\kappa_{21}\kappa_{12}\kappa_{22},\\
&&e^{\ga_{31}}=-\al_{1}^{2}\al_{2}\ba_{1}^{2}\ba_{2}^{2}\varrho_{12}^2\bar{\varrho}_{12}\kappa_{11}^2\kappa_{21}^2\kappa_{12}\kappa_{22},~e^{\ga_{32}}=-\al_{1}\al_{2}^{2}\ba_{1}^{2}\ba_{2}^{2}\varrho_{12}^2\bar{\varrho}_{12}\kappa_{11}\kappa_{21}\kappa_{12}^2\kappa_{22}^2.
\eea 
The auxiliary functions involved in the derivation of two soliton solution are given below,
\bes
\bea
s^{(1)}(-x,t)&=&\al_{1}^{2}e^{2\bar{\xi}_1}+ \al_{2}^{2}e^{2\bar{\xi}_2}+2\al_{1}\al_{2}e^{\bar{\xi}_1+\bar{\xi}_2}+e^{\xi_1+2\bar{\xi}_1+\bar{\xi}_2+\phi_1}+e^{2\bar{\xi}_1+\xi_2+\bar{\xi}_2+\phi_2} \nonumber \\ &&
+e^{\xi_1+\bar{\xi}_1+2\bar{\xi}_2+\phi_3}+e^{\bar{\xi}_1+\xi_2+2\bar{\xi}_2+\phi_4}+e^{2(\xi_1+\bar{\xi}_1+\bar{\xi}_2)+\phi_{11}}\nonumber \\&&
+e^{2(\xi_2+\bar{\xi}_1+\bar{\xi}_2)+\phi_{12}} +e^{\xi_1+2\bar{\xi}_1+\xi_2+2\bar{\xi}_2+\phi_{13}},\\
s^{(2)}(-x,t)&=&\ba_{1}^{2}e^{2\xi_1}+ \ba_{2}^{2}e^{2\xi_2}+2\ba_{1}\ba_{2}e^{\xi_1+\xi_2}+e^{2\xi_1+\bar{\xi}_1+\xi_2+\psi_1}+e^{\bar{\xi}_2+\xi_2+2\xi_1+\psi_2} \nonumber \\ 
&&+e^{\xi_1+\bar{\xi}_1+2\xi_2+\psi_3}+e^{\xi_1+2\xi_2+\bar{\xi}_2+\psi_4}+e^{2(\xi_1+\bar{\xi}_1+\xi_2)+\psi_{11}} \nonumber \\
&&+e^{2(\xi_2+\xi_1+\bar{\xi}_2)+\psi_{12}}+e^{2\xi_1+\bar{\xi}_1+2\xi_2+\bar{\xi}_2+\psi_{13}},\eea
\bea
h^{(1)}(-x,t)&=&\al_1\ba_1e^{\xi_1+\bar{\xi}_1}+\al_1\ba_2e^{\xi_2+\bar{\xi}_1}+\al_2\ba_1e^{\xi_1+\bar{\xi}_2}+\al_2\ba_2e^{\xi_2+\bar{\xi}_2}\nonumber\\&&
+e^{2\xi_1+\bar{\xi}_1+\bar{\xi}_2+\varphi_1} +e^{2\xi_2+\bar{\xi}_1+\bar{\xi}_2+\varphi_2}+e^{2\bar{\xi}_1+\xi_1+\xi_2+\varphi_3}+e^{2\bar{\xi}_2+\xi_1+\xi_2+\varphi_4}\nonumber\\&&
+e^{\bar{\xi}_1+\bar{\xi}_2+\xi_1+\xi_2+\varphi_5}+e^{2\xi_1+2\bar{\xi}_1+\xi_2+\bar{\xi}_2+\varphi_{11}}+e^{\xi_1+2\bar{\xi}_1+2\xi_2+\bar{\xi}_2+\varphi_{12}}\nonumber\\&&
+e^{2\xi_1+\bar{\xi}_1+\xi_2+2\bar{\xi}_2+\varphi_{13}}+e^{\xi_1+\bar{\xi}_1+2\xi_2+2\bar{\xi}_2+\varphi_{14}}=h^{(2)}(-x,t),
\eea \ees
\bea
&&\nonumber e^{\phi_1}=-2\bar{\varrho}_{12}\al_1^{2}\al_2\ba_1\kappa_{11}\kappa_{12},~ e^{\phi_2}=-2\bar{\varrho}_{12}\al_1^{2}\al_2\ba_2\kappa_{21}\kappa_{22},\\
&&\nonumber e^{\phi_3}=-2\bar{\varrho}_{12}\al_1\al_2^{2}\ba_1\kappa_{11}\kappa_{12},~ e^{\phi_4}=-2\bar{\varrho}_{12}\al_1\al_2^{2}\ba_2\kappa_{21}\kappa_{22},\\
&&\nonumber e^{\phi_{11}}=\bar{\varrho}_{12}^2\al_1^{2}\al_2^2\ba_1^2\kappa_{11}^2\kappa_{12}^2,~  e^{\phi_{12}}=\bar{\varrho}_{12}^2\al_1^{2}\al_2^2\ba_2^2\kappa_{21}^2\kappa_{22}^2,\\
&&\nonumber e^{\phi_{13}}=2\bar{\varrho}_{12}^2\al_1^{2}\al_2^2\ba_1\ba_2\kappa_{11}\kappa_{12}\kappa_{21}\kappa_{22},~ e^{\psi_1}=-2\varrho_{12}\al_1\ba_1^2\ba_2\kappa_{11}\kappa_{21},\\
&&\nonumber e^{\psi_2}=-2\varrho_{12}\al_2\ba_1^2\ba_2\kappa_{12}\kappa_{22},~ e^{\psi_3}=-2\varrho_{12}\al_1\ba_1\ba_2^2
\kappa_{11}\kappa_{21},\\
&&\nonumber e^{\psi_4}=-2\varrho_{12}\al_2\ba_1\ba_2^2\kappa_{12}\kappa_{22},~ e^{\psi_{11}}=\varrho_{12}^2\al_2^2\ba_1^2\ba_2^2\kappa_{11}^2\kappa_{21}^2,~ e^{\psi_{12}}=\varrho_{12}^2\al_2^2\ba_1^2\ba_2^2\kappa_{12}^2\kappa_{22}^2,\\
&&\nonumber e^{\psi_{13}}=2\varrho_{12}^2\al_1\al_2\ba_1^2\ba_2^2\kappa_{11}\kappa_{21}\kappa_{12}\kappa_{22},~ e^{\varphi_1}=-\bar{\varrho}_{12}\al_1\al_2\ba_1^2\kappa_{11}\kappa_{12},\\
&&\nonumber e^{\varphi_2}=-\bar{\varrho}_{12}\al_1\al_2\ba_2^2\kappa_{21}\kappa_{22},~ e^{\varphi_3}=-\varrho_{12}\al_1^2\ba_1\ba_2\kappa_{11}\kappa_{21},~ e^{\varphi_4}=-\varrho_{12}\al_2^2\ba_1\ba_2\kappa_{12}\kappa_{22},\\
&&\nonumber e^{\varphi_5}=-2\al_1\al_2\ba_1\ba_2\kappa_{11}\kappa_{21}\kappa_{12}\kappa_{22}\nu_2,\\
&&\nu_2=\{-\bar{k}_1k_2\bar{k}_2(k_2-\bar{k}_2)(k_2+\bar{k}_2)^2+k_2^2\bar{k}_2^2(k_2+\bar{k}_2)^2+\bar{k}_1^4(k_2^2+k_2\bar{k}_2+\bar{k}_2^2)\nonumber\\
&&\hspace{1.0cm}+k_1^4(\bar{k}_1^2+k_2^2+\bar{k}_1(k_2-\bar{k}_2)+k_2\bar{k}_2+\bar{k}_2^2)+\bar{k}_1^3(2k_2^3+\bar{k}_2k_2^2-k_2\bar{k}_2^2 \nonumber\\
&&\hspace{1.0cm}-2\bar{k}_2^3)+\bar{k}_1^2(k_2^4-k_2^3\bar{k}_2-3k_2^2\bar{k}_2^2-k_2\bar{k}_2^3+\bar{k}_2^4)+k_1^3(2\bar{k}_1^3-2k_2^3+\bar{k}_1^2(k_2\nonumber \\
&&\hspace{1.0cm}-\bar{k}_2)-k_2^2\bar{k}_2+k_2\bar{k}_2^2+2\bar{k}_2^3-\bar{k}_1(k_2^2+\bar{k}_2^2))+k_1^2(\bar{k}_1^{4}+k_2^4-k_2^3\bar{k}_2 \nonumber \\
&&\hspace{1.0cm}-3k_2^2\bar{k}_2^2-k_2\bar{k}_2^3+\bar{k}_2^4+\bar{k}_1^3(\bar{k}_2-k_2)-3\bar{k}_1^2(k_2^2+\bar{k}_2^2)+\bar{k}_1(\bar{k}_2^3-k_2^3))\nonumber \\
&&\hspace{1.0cm}+k_1(\bar{k}_1^4(-k_2+\bar{k}_2)+k_2\bar{k}_2(k_2-\bar{k}_2)(k_2+\bar{k}_2)^2-\bar{k}_1^3(k_2^2+\bar{k}_2^2)\nonumber\\
&&\hspace{1.0cm}+\bar{k}_1^2(k_2^3-\bar{k}_2^3)+\bar{k}_1(k_2^4+\bar{k}_2^4))\}\nonumber,\\
&&\nonumber e^{\varphi_{11}}=\bar{\varrho}_{12}\varrho_{12}\al_1^2\al_2\ba_1^2\ba_2\kappa_{11}^2\kappa_{21}\kappa_{12},~ e^{\varphi_{12}}=\bar{\varrho}_{12}\varrho_{12}\al_1^2\al_2\ba_1\ba_2^2\kappa_{11}\kappa_{21}^2\kappa_{22},\\
&&e^{\varphi_{13}}=\bar{\varrho}_{12}\varrho_{12}\al_1\al_2^2\ba_1^2\ba_2\kappa_{11}\kappa_{12}^2\kappa_{22},~ e^{\varphi_{14}}=\bar{\varrho}_{12}\varrho_{12}\al_1\al_2^2\ba_1\ba_2^2\kappa_{21}\kappa_{12}\kappa_{22}^2.
\eea 
where $\bar{\varrho}_{12}=(\bar{k}_1-\bar{k}_2)^2$, $\varrho_{12}=(k_1-k_2)^2$, $\kappa_{lm}=\frac{1}{(k_l+\bar{k}_m)^2}$, $l,m=1,2$.
\section*{References}

\bibliography{mybibfile}

\end{document}